\documentclass[aps,prl,twocolumn,showpacs,superscriptaddress]{revtex4}
\usepackage{graphicx}
\usepackage{dcolumn}
\usepackage{bm}
\usepackage{color}

\newcommand{\sqrts}{\mbox{$\sqrt{s}$}}
\newcommand{\sqrtsNN}{\mbox{$\sqrt{s_{_{\mathrm{NN}}}}$}}
\newcommand{\dAu}{\textit{d}+Au}

\newcommand{\AuAu}{Au+Au}

\renewcommand{\AA}{\mbox{A+A}}

\newcommand{\pp}{\mbox{\textit{p}+\textit{p}}}
\newcommand{\eeh}{\mbox{$(e^{+}$+$e^{-})/2$}}

\newcommand{\pt}{\mbox{$p_T$}}

\newcommand{\gevcc}{\mbox{$\mathrm{GeV/}c^2$}}
\newcommand{\mevcc}{\mbox{$\mathrm{MeV/}c^2$}}
\newcommand{\gevc}{\mbox{${\mathrm{GeV/}}c$}}

\newcommand{\RAA}{\mbox{$R_{AA}$}}

\newcommand{\dedx}{\mbox{$dE/dx$}}

\begin{document}

\title{Transverse momentum and centrality dependence of high-\pt\
    non-photonic electron suppression in Au+Au collisions at \sqrtsNN\
    = 200 GeV}

\date{\today}

\affiliation{Argonne National Laboratory, Argonne, Illinois 60439}
\affiliation{University of Birmingham, Birmingham, United Kingdom}
\affiliation{Brookhaven National Laboratory, Upton, New York 11973}
\affiliation{California Institute of Technology, Pasadena,
California 91125} \affiliation{University of California, Berkeley,
California 94720} \affiliation{University of California, Davis,
California 95616} \affiliation{University of California, Los
Angeles, California 90095} \affiliation{Carnegie Mellon University,
Pittsburgh, Pennsylvania 15213} \affiliation{University of Illinois,
Chicago} \affiliation{Creighton University, Omaha, Nebraska 68178}
\affiliation{Nuclear Physics Institute AS CR, 250 68
\v{R}e\v{z}/Prague, Czech Republic} \affiliation{Laboratory for High
Energy (JINR), Dubna, Russia} \affiliation{Particle Physics
Laboratory (JINR), Dubna, Russia} \affiliation{University of
Frankfurt, Frankfurt, Germany} \affiliation{Institute of Physics,
Bhubaneswar 751005, India} \affiliation{Indian Institute of
Technology, Mumbai, India} \affiliation{Indiana University,
Bloomington, Indiana 47408} \affiliation{Institut de Recherches
Subatomiques, Strasbourg, France} \affiliation{University of Jammu,
Jammu 180001, India} \affiliation{Kent State University, Kent, Ohio
44242} \affiliation{Institute of Modern Physics, Lanzhou, China}
\affiliation{Lawrence Berkeley National Laboratory, Berkeley,
California 94720} \affiliation{Massachusetts Institute of
Technology, Cambridge, MA 02139-4307}
\affiliation{Max-Planck-Institut f\"ur Physik, Munich, Germany}
\affiliation{Michigan State University, East Lansing, Michigan
48824} \affiliation{Moscow Engineering Physics Institute, Moscow
Russia} \affiliation{City College of New York, New York City, New
York 10031} \affiliation{NIKHEF and Utrecht University, Amsterdam,
The Netherlands} \affiliation{Ohio State University, Columbus, Ohio
43210} \affiliation{Panjab University, Chandigarh 160014, India}
\affiliation{Pennsylvania State University, University Park,
Pennsylvania 16802} \affiliation{Institute of High Energy Physics,
Protvino, Russia} \affiliation{Purdue University, West Lafayette,
Indiana 47907} \affiliation{Pusan National University, Pusan,
Republic of Korea} \affiliation{University of Rajasthan, Jaipur
302004, India} \affiliation{Rice University, Houston, Texas 77251}
\affiliation{Universidade de Sao Paulo, Sao Paulo, Brazil}
\affiliation{University of Science \& Technology of China, Hefei
230026, China} \affiliation{Shanghai Institute of Applied Physics,
Shanghai 201800, China} \affiliation{SUBATECH, Nantes, France}
\affiliation{Texas A\&M University, College Station, Texas 77843}
\affiliation{University of Texas, Austin, Texas 78712}
\affiliation{Tsinghua University, Beijing 100084, China}
\affiliation{Valparaiso University, Valparaiso, Indiana 46383}
\affiliation{Variable Energy Cyclotron Centre, Kolkata 700064,
India} \affiliation{Warsaw University of Technology, Warsaw, Poland}
\affiliation{University of Washington, Seattle, Washington 98195}
\affiliation{Wayne State University, Detroit, Michigan 48201}
\affiliation{Institute of Particle Physics, CCNU (HZNU), Wuhan
430079, China} \affiliation{Yale University, New Haven, Connecticut
06520} \affiliation{University of Zagreb, Zagreb, HR-10002, Croatia}

\author{B.I.~Abelev}\affiliation{University of Illinois, Chicago}
\author{M.M.~Aggarwal}\affiliation{Panjab University, Chandigarh 160014, India}
\author{Z.~Ahammed}\affiliation{Variable Energy Cyclotron Centre, Kolkata 700064, India}
\author{B.D.~Anderson}\affiliation{Kent State University, Kent, Ohio 44242}
\author{D.~Arkhipkin}\affiliation{Particle Physics Laboratory (JINR), Dubna, Russia}
\author{G.S.~Averichev}\affiliation{Laboratory for High Energy (JINR), Dubna, Russia}
\author{Y.~Bai}\affiliation{NIKHEF and Utrecht University, Amsterdam, The Netherlands}
\author{J.~Balewski}\affiliation{Indiana University, Bloomington, Indiana 47408}
\author{O.~Barannikova}\affiliation{University of Illinois, Chicago}
\author{L.S.~Barnby}\affiliation{University of Birmingham, Birmingham, United Kingdom}
% ...

\collaboration{STAR Collaboration}\noaffiliation

\begin{abstract}

The STAR collaboration at RHIC reports measurements of the inclusive
yield of non-photonic electrons, which arise dominantly from
semi-leptonic decays of heavy flavor mesons, over a broad range of
transverse momenta ($1.2 < \pt < 10$ \gevc) in \pp, \dAu, and \AuAu\
collisions at \sqrtsNN\ = 200 GeV. The non-photonic electron yield
exhibits unexpectedly large suppression in central \AuAu\ collisions
at high \pt, suggesting substantial heavy quark energy loss at RHIC.
The centrality and \pt\ dependences of the suppression provide
constraints on theoretical models of suppression.

\end{abstract}
\maketitle


High \pt\ hadron production measurements at the Relativistic Heavy Ion
Collider (RHIC) show a strong suppression of the single-particle
inclusive yields in nuclear collisions
\cite{star130auau,star200auau,star200dau}. The suppression is commonly
thought to arise from partonic energy loss in dense matter due to
induced gluon radiation \cite{gluonrad}, with its magnitude depending
strongly on the color charge density of the medium. This makes it a
sensitive probe of the matter created in heavy-ion collisions, where a
quark-gluon plasma may form if sufficient energy density is achieved.

Charm and bottom quarks are produced dominantly through high-$Q^{2}$
partonic interactions. Heavy flavor cross-sections and \pt\ spectra have
been calculated at next-to-leading-order (NLO) for both \pp\ and
\AA\ collisions \cite{pQCD, pQCDCalc1, pQCDCalc2}, including nuclear matter effects \cite{pQCDCalc2}.
Although pQCD calculations agree well with heavy
quark production in collider experiments at higher \sqrts\
\cite{Frixione:2005yf}, they disagree with
recent RHIC measurements \cite{STARDMesons1, PhenixPP}.  Nevertheless,
measurements of heavy quark production potentially provide new
constraints on partonic energy loss mechanisms \cite{dead, dead1,
elastic, Djordjevic:2005, Armesto, Wicks:2005, Hess:2005}.  Gluon
radiation in a forward cone is suppressed for heavy quarks at moderate
energy (dead cone effect) \cite{dead,dead1}, with corresponding
reduction in medium induced energy loss and less suppression of
heavy-quark mesons than light quark mesons.


Direct reconstruction of heavy flavor mesons via
hadronic decay channels \cite{STARDMesons1} is difficult in the
complex environment of high energy nuclear collisions. Heavy quark
production can also be studied through measurements of electrons
(positrons) from semileptonic $D$ and $B$ decays. This Letter reports
STAR Collaboration measurements of the non-photonic electron
yield, \eeh, in \pp, \dAu, and \AuAu\ collisions at nucleon-nucleon
center of mass energy \sqrtsNN\ = 200 GeV. The data extend
significantly the \pt\ range of previous electron suppression studies
\cite{Adler:2005xv}, to a region of phase-space where bottom decays
are expected to be dominant.  Large differences in energy loss are
expected between $c$ and $b$ quarks in this region
\cite{Djordjevic:2005}, and these measurements provide important new
constraints on partonic energy loss mechanisms.


STAR is a large acceptance apparatus comprising several detector
subsystems within a 0.5 T solenoidal magnet field
\cite{Ackermann:2002ad}. The main detectors for this analysis are
the Time Projection Chamber (TPC) \cite{refTPC} and the barrel
Electromagnetic Calorimeter (EMC) \cite{Beddo:2002zx}.  The EMC has
a gas-filled Shower Maximum Detector (SMD) at a depth of $\sim5
X_{0}$ to measure shower shape and position. A fast trigger based on
single EMC tower energy enriches the electron sample at high \pt.
Electrons at moderate \pt\ were reconstructed from minimum bias and
centrality triggered Au+Au event samples, while EMC triggered events
were used for $\pt> 3 - 4$ \gevc. \AuAu\ data were divided into 3
centrality classes based on the track multiplicity measured at
midrapidity. The integrated luminosity sampled by the EMC trigger is
100 nb$^{-1}$ for \pp, 370 $\mu$b$^{-1}$ for \dAu\, and 26
$\mu$b$^{-1}$ for the most central \AuAu\ events.  The charged
particle acceptance is $0<\eta<0.7$ and $0<\phi<2\pi$, selected to
minimize the radiation length of detector material interior to the
EMC within the available EMC acceptance.


The analysis has three main steps: selection of
electrons; subtraction of background from
decays and interactions in material; and residual corrections to the
signal yield. Table \ref{table1} shows the major correction factors
and uncertainties, which we now discuss in detail.

\begin{figure}[tbh]
    \includegraphics[width=0.46\textwidth]{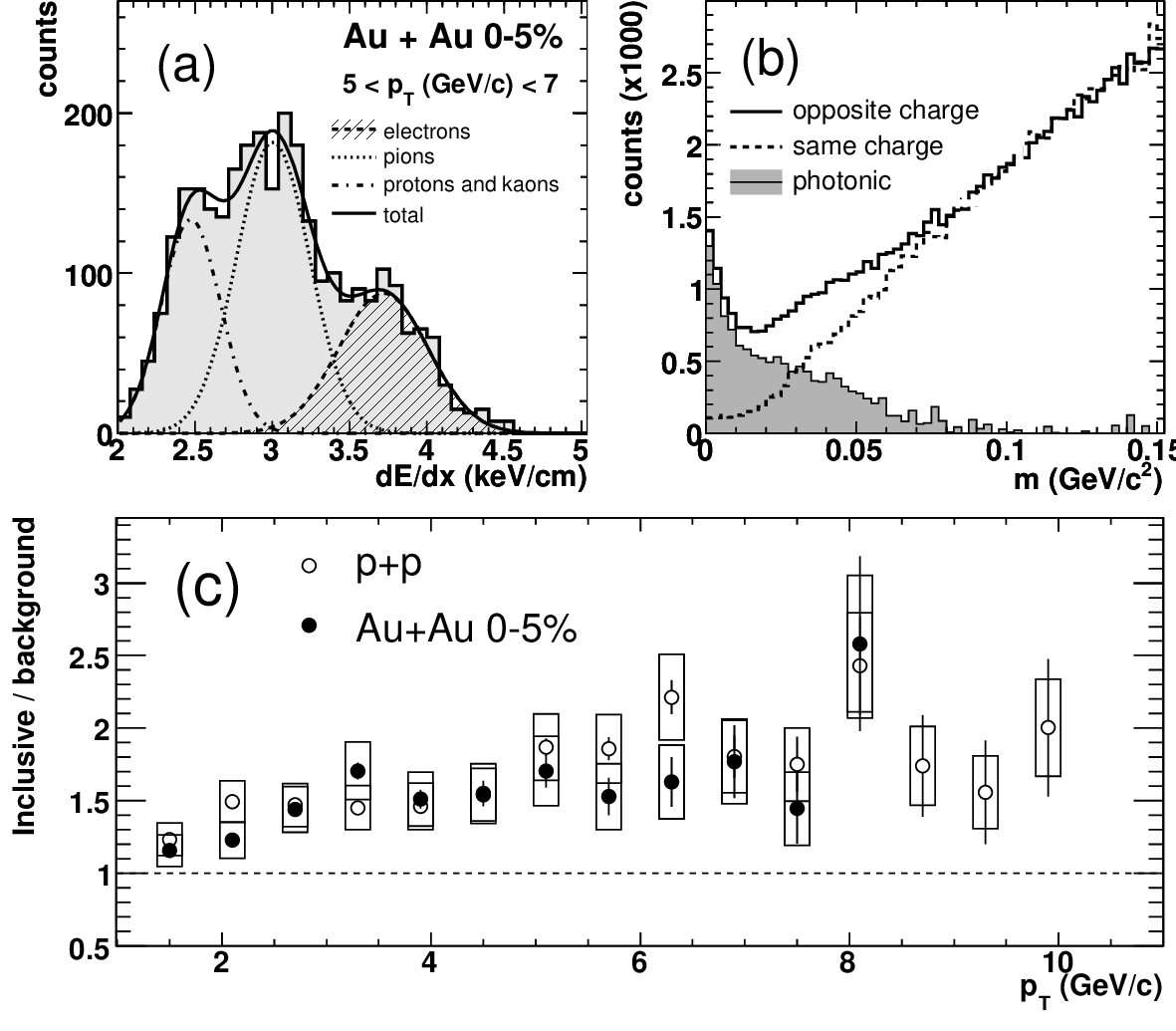} \caption{ (a)
    \dedx\ projections for $5 < \pt (\gevc) < 7$ in central \AuAu\
    events after EMC and SMD cuts. The lines are Gaussian fits for
    $p+K$, $\pi$, and electron yields. (b) Invariant $e^{+}e^{-}$ mass
    spectrum.  (c) Ratio of inclusive and background electron yield
    vs. \pt\ for \pp\ and \AuAu\ collisions. Vertical bars are
    statistical errors, boxes are systematic uncertainties.}
    \label{fig01}
\end{figure}

{\it Electron PID:} Electron identification utilizes ionization
energy loss (\dedx) and track momentum from the TPC, together with
energy and shower shapes from the EMC. Tracks with momentum $p>1.5$
\gevc\ are accepted if they originate from the primary vertex
(distance of closest approach less than 1.5 cm) and project to an
active EMC tower, with acceptance $\alpha_{\mathrm{EMC}}\sim75-85\%$
of the EMC instrumented coverage. This reduced acceptance is due to
dead or noisy electronics channels. Initial electron identification
is based on $p/E<2$, where $p$ is the TPC track momentum and $E$ is
the energy of the EMC tower. Simulations show that this cut excludes
$\sim$ 7\% of real electrons due to sharing of shower energy between
towers. Additional hadron rejection is based on the shower shape
measured by the SMD. Figure \ref{fig01}a shows the \dedx\
distribution for tracks passing the $p/E$ and shower shape cuts. The
curves show Gaussian functions fit to the distribution, representing
the yields of $p+K$, pions and electrons \cite{Shao}. The parameters
in the fit are the yields, widths, and overall \dedx\ scale, with
widths and the distances between centroids being quasi-free
parameters, constrained by a model of energy deposition in the TPC
gas \cite{Bichsel}.

Electrons are selected by cutting on TPC energy loss
$\dedx_{min}<\dedx<5.1$ keV/cm. $\dedx_{min}$ is around 3.5 keV/cm,
with the specific value having weak dependence on the event
multiplicity and increasing slowly with track momentum, to optimize
electron efficiency and hadron rejection while preserving more than
50\% of the electrons in the \dedx\ distribution. The residual
hadron background satisfying the $\dedx$ cut is estimated based on
Gaussian fits similar to those in Figure \ref{fig01}.

\begingroup
\squeezetable
\begin{table} [tb]
\caption{Corrections and systematic uncertainties for
the non-photonic electron yield at \pt\ = 2 and 8 GeV/c.}
\begin{ruledtabular}
\begin{tabular}{lcccc}
Correction                 &   \multicolumn{2}{c}{\pp}            &  \multicolumn{2}{c}{central \AuAu} \\
                           &   2 \gevc\         & 8 \gevc\        &  2 \gevc\ &      8 \gevc\  \\ \hline
Acceptance                 &   \multicolumn{2}{c}{0.84$\pm$0.05}  & \multicolumn{2}{c}{0.75$\pm$0.15}  \\
PID efficiency             &   0.25$\pm$0.03  & 0.50$\pm$0.03     & 0.13$\pm$0.03    & 0.45$\pm$0.03 \\
hadron contamination       &   $<$0.01           & 0.20$\pm$0.04    & 0.03$\pm$0.03        & 0.22$\pm$0.05 \\
Bkgd.~reco.~eff.~($\varepsilon_{B}$)           &   0.65$\pm$0.06 & 0.55$\pm$0.06 & 0.56$\pm$0.06 &  0.50$\pm$0.06 \\
Bremsstr.~\& $\delta p/p$  & 0.86$\pm$0.14    & 1.05$\pm$0.05     &  0.9$\pm$0.1  &     1.1$\pm$0.1  \\
EMC $\varepsilon_{\mathrm{trigger}}$ &  --   &  1.00$\pm$0.08     & -- & 1.00$\pm$0.05 \\
Cross section              &  \multicolumn{2}{c}{$\pm$0.14}       &  \multicolumn{2}{c}{--} \\
\end{tabular}
\end{ruledtabular}
\label{table1}
\end{table}
\endgroup

Table \ref{table1} shows the combined electron tracking and
identification efficiency (``PID efficiency''), determined by
embedding simulated electrons into real events. It is significantly
below unity due to tracking efficiency ($\sim70$\%), exclusion of
electrons due to the energy leakage to neighboring towers, and SMD
response. Its increase from $\pt=2$ to 8 \gevc\ is due to increasing
SMD efficiency.

{\it Electron background:} Background from photonic sources is due
largely to photon conversions ($\sim85$\%) in the detector material
between the interaction point and the TPC ($X/X_{0} \sim 4.5\%$) and
$\pi^{0}$ and $\eta$ Dalitz decays \cite{Dalitz} ($\sim15$\%). The
photonic electron yield is measured using the invariant mass
distribution of track pairs detected in the TPC. One track of the
pair is required to fall in the EMC acceptance, satisfying $p>1.5$
\gevc\ and electron PID cuts, with the other track having $\pt>0.15$
\gevc\ within the TPC acceptance and a loose cut around the electron
$dE/dx$ band. Figure~\ref{fig01}b shows the invariant mass
distribution of pairs with the same or opposite charge sign. The
same-sign distribution is due to random (combinatorial) pairs. An
alternative combinatorial distribution formed by embedding single
simulated electrons into real events agrees with the same-sign
distribution within statistical uncertainties.

The shaded region in Figure~\ref{fig01}b is the difference between
the opposite and same-sign distributions and represents the photonic
yield. It exhibits a peak at zero invariant mass due to conversions,
and a tail at non-zero mass due to Dalitz decays \cite{Dalitz}.
Selecting $m<150$ \mevcc\ accepts $\sim98$\% of all $\pi^0$ and
$\eta$ Dalitz pairs in this distribution. The efficiency
$\varepsilon_{B}(\pt)$ to identify a photonic electron in the EMC by
this procedure was estimated by embedding \cite{embedding} the main
background sources ($\pi^{0}$ and $\gamma$) with a realistic
momentum distribution derived from recent RHIC data \cite{PHENIXpi0}
into real events.

The photonic electron yield $N_{ph}$ is calculated in each \pt\ bin
via $N_{ph} = (N_{unlike}-N_{like})/\varepsilon_{B}$.  Additional
background, mainly from $\omega$, $\phi$, and $\rho$ decays, was
estimated using PYTHIA \cite{Sjostrand:2003wg} and HIJING
\cite{Gyulassy:1994ew} simulations to be $\sim 2-4\%$ of $N_{ph}$
\cite{STARDMesons1} and is included in the systematic uncertainty of
$N_{ph}$.  Figure \ref{fig01}c depicts the ratio of the inclusive to
the photonic electron spectra for \pp\ and \AuAu\ collisions. The
Figure shows a clear electron excess. Within uncertainties, the
non-photonic excess is independent of centrality at high \pt.

{\it Non-photonic electron yield:} The trigger efficiency was
determined by comparing the electron candidate spectrum in the
minimum bias and triggered data sets. At high-\pt\ the ratio of the
spectra is compatible with the online scale-down factor applied to
minimum bias events. The non-photonic spectrum is the difference of
the inclusive and photonic spectra.  Additional corrections are
applied for momentum resolution and bremsstrahlung, determined from
simulations.

{\it Systematic uncertainties:}
Systematic uncertainties were determined by varying cut parameters
within reasonable limits. The uncorrelated systematic uncertainty of the electron
yield is dominated by the electron identification efficiency and
photonic background reconstruction at low \pt\ and the correction
for residual hadron background at high \pt.


\begin{figure}[tb]
    \includegraphics[width=0.46\textwidth]{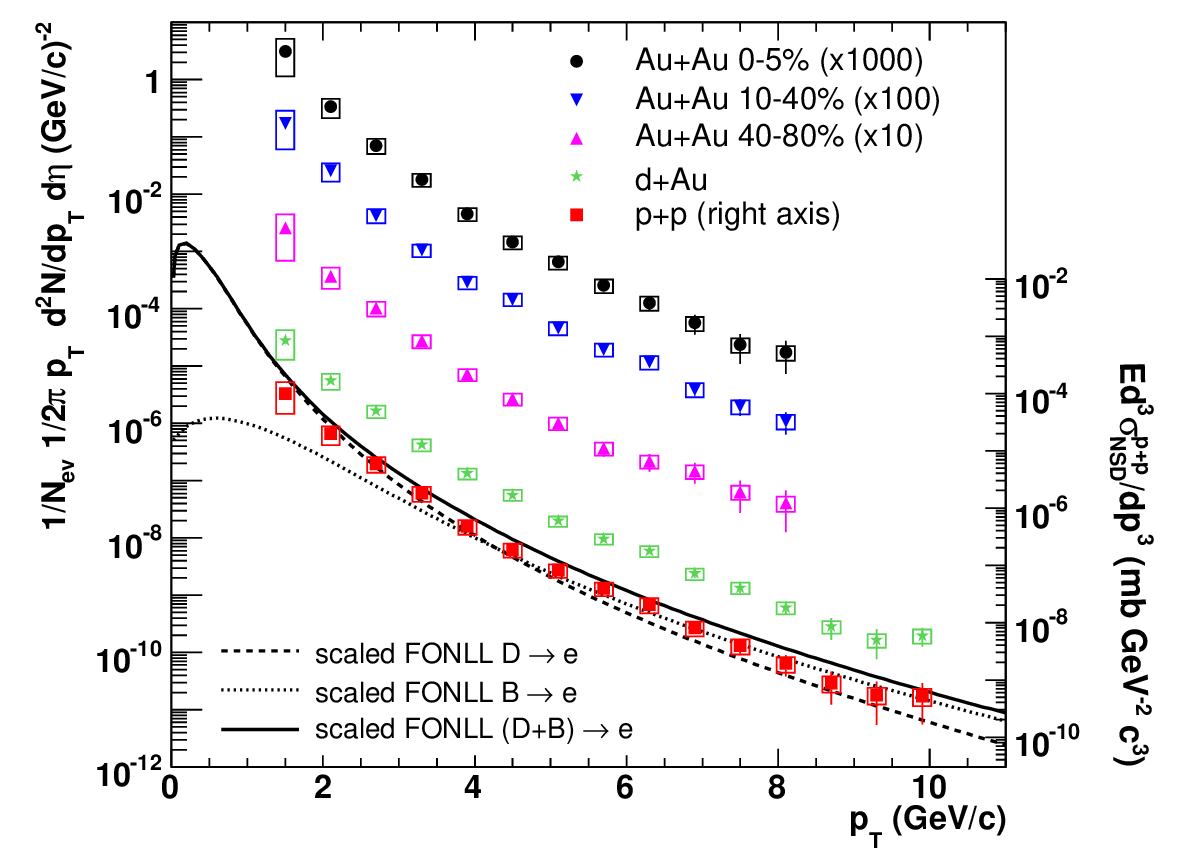}
    \caption{Non-photonic electron spectra. Vertical bars are
    statistical errors, boxes are systematic uncertainties. The curves
    are scaled pQCD predictions for \pp\ \cite{pQCDCalc2}. Cross
    section on right axis applies to \pp\ spectrum only.}
    \label{fig02}
\end{figure}

Figure \ref{fig02} shows the fully corrected non-photonic electron
spectra for 200 GeV \pp, \dAu, and \AuAu\ collisions. The curves correspond to
FONLL (Fixed Order Next-to-Leading Log) predictions
\cite{pQCDCalc2} for semileptonic $D$ and
$B$ meson decays. The calculated spectrum is scaled by 5.5 (see
below).

Figure \ref{fig03}, upper part (points), shows the ratio of measured
to unscaled FONLL-calculated non-photonic electron yield for \pp\
collisions. The calculation describes the \textit{shape} of the
measured spectra relatively well, though with a large difference in
their overall scale. Better agreement is found at larger $\sqrt{s}$
\cite{Frixione:2005yf}. The same ratio is shown for published STAR
\cite{STARDMesons1} and PHENIX \cite{PhenixPP} measurements. The
horizontal dashed line is at 5.5 $\pm$ 0.8(\emph{stat}) $\pm$
1.7(\emph{sys}), corresponding to the ratio between the total charm
cross section measured by STAR \cite{STARDMesons1} to the central
value predicted by FONLL \cite{pQCDCalc2,Frixione:2005yf}. The
shaded band around that line shows the experimental uncertainty in
this ratio. PHENIX data \cite{PhenixPP} exhibit a lower ratio and
appear not to be consistent with the data reported here. The lower
part (curves) shows the relative contribution to the FONLL
calculation of charm and bottom decays, with the variation due to
NLO uncertainties \cite{pQCDCalc2, RAMONA}. The $B$-decay
contribution is expected to be significant in the upper \pt\ range
of this measurement.

Modification of the inclusive particle production is measured by the nuclear modification factor
\cite{star130auau} ($\RAA(\pt)$). \RAA\ is unity for hard processes without nuclear effects. Figure \ref{fig04} shows \RAA(\pt)
for non-photonic electrons in
\dAu\ and central \AuAu\ collisions. Error bars show the statistical
uncertainties, boxes show uncorrelated systematic uncertainties, and
the filled band at unity is the overall normalization uncertainty. \RAA\ for
\dAu\ is consistent with a moderate Cronin enhancement. \RAA$\sim0.2$ for central \AuAu\
collisions at $\pt>3$ \gevc, consistent with a previous measurement at
lower \pt\
\cite{Adler:2005xv}. The suppression is similar to that for light
hadrons at $\pt\ > 6$ \gevc\ \cite{star200auau}.

\begin{figure}[tb]
    \includegraphics[width=0.46\textwidth]{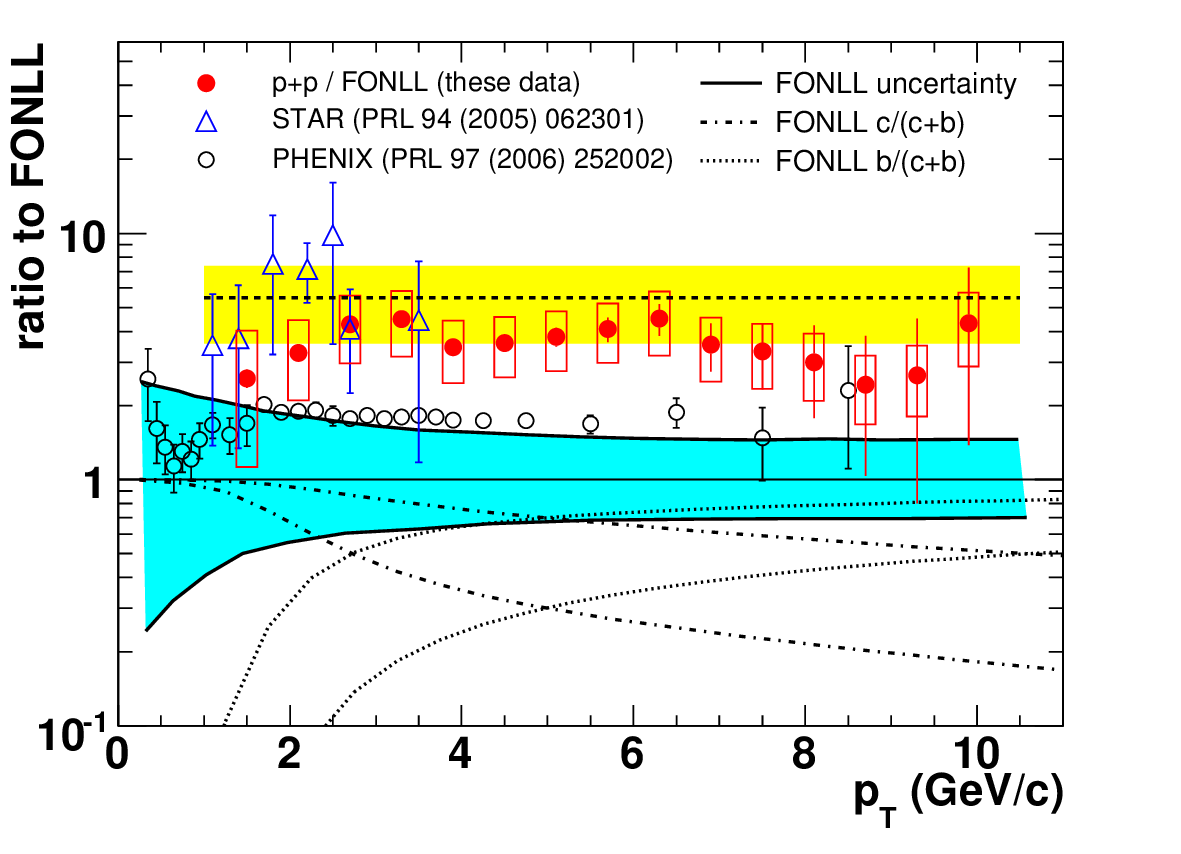}
    \caption{Upper: ratio between measured non-photonic electron yield
        and FONLL pQCD calculations \cite{pQCDCalc2} for \pp\
        collisions. Lower: relative contributions to FONLL distribution
        of c and b decays. }
    \label{fig03}
\end{figure}


Figure \ref{fig04} shows predictions for electron \RAA\ from
semi-leptonic D- and B-meson decay in central \AuAu\ collisions
using calculations of heavy quark energy loss. Curve I uses DGLV
radiative energy loss via few hard scatterings
\cite{Djordjevic:2005} with initial gluon density $dN_{g}/dy=1000$,
consistent with light quark suppression. Curve II uses BDMPS
radiative energy loss via multiple soft collisions \cite{Armesto},
with transport coefficient $\hat{q}$. $\hat{q}$ is set to $14$
GeV$^2$/fm, though light quark hadron suppression provides only a
loose constraint $4 < \hat{q}<14$ GeV$^{2}$/fm \cite{Armesto}. Both
calculations predict much less suppression than observed.

\begin{figure}[t]
    \includegraphics[width=0.46\textwidth]{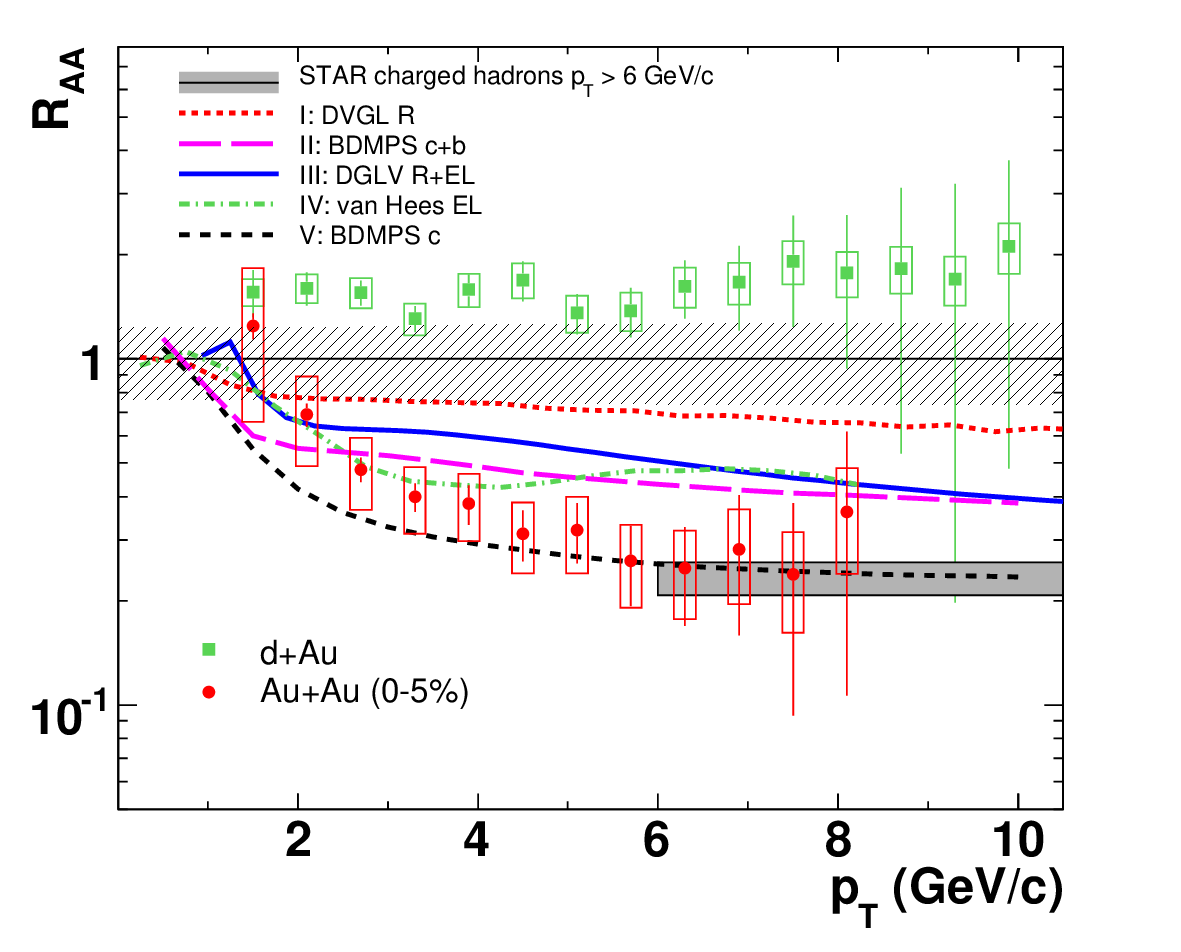} \caption{The
    nuclear modification factor, \RAA, for \dAu\ and \AuAu\ collisions
    at \sqrtsNN = 200 GeV. Error bars and uncertainties are described
    in text.}  \label{fig04}
\end{figure}

This discrepancy may indicate significant collisional (elastic)
energy loss for heavy quarks \cite{elastic, Magdalena}. Curve III is a
DGLV-based calculation including both radiative \textit{and}
collisional energy loss, together with path length fluctuations
\cite{Wicks:2005}. The calculated
suppression is also markedly less than that observed. For
Curve IV, the heavy quark energy loss is due to elastic scattering
mediated by resonance excitations ($D$ and $B$) and LO t-channel gluon
exchange \cite{Hess:2005}. This calculation also predicts
significantly less suppression than observed.

Dead cone reduction of energy loss is expected to be more
significant for bottom than charm quarks in the reported \pt\ range.
Curve V, which is the same calculation as curve II but for D-meson
decays only, agrees better with the data. Since there is better
agreement of data and theory for bottom than charm production at the
Tevatron \cite{Frixione:2005yf}, the scale factor 5.5 between
calculated and measured \pp\ electron yields may overestimate the
$B$ decay contribution at RHIC, i.e. $D$ decays may in fact dominate
the electron yields in the reported \pt\ range, favoring calculation
V. A direct measurement of D-mesons at high-\pt\ is required to
understand energy loss of heavy quarks in detail. Finally,
multi-body mechanisms may also contribute to heavy quark energy loss
\cite{Liu}.


We have reported the measurement of high-\pt\ non-photonic electrons
in \pp, \dAu, and \AuAu\ collisions at \sqrtsNN\ = 200 GeV.
A pQCD calculation for heavy quark production in \pp\ collisions underpredicts
the data, although it describes the overall shape of the \pt\ distribution
relatively well.
Large yield suppression is observed in central
\AuAu\ collisions, consistent with substantial energy loss of heavy
quarks in dense matter. The suppression is larger than that expected
from radiative energy loss calculations, suggesting that other
processes contribute significantly to heavy quark energy
loss. This unique sensitivity to the energy loss mechanisms makes
the measurement of heavy quark suppression an essential component of the
study of dense matter. Full description of the interaction between
partons and the medium will require further detailed measurements of
charm and bottom separately.


We thank the RHIC Operations Group and RCF at BNL, and the
NERSC Center at LBNL for their support. This work was supported
in part by the HENP Divisions of the Office of Science of the U.S.
DOE; the U.S. NSF; the BMBF of Germany; IN2P3, RA, RPL, and
EMN of France; EPSRC of the United Kingdom; FAPESP of Brazil;
the Russian Ministry of Science and Technology; the Ministry of
Education and the NNSFC of China; IRP and GA of the Czech Republic,
FOM of the Netherlands, DAE, DST, and CSIR of the Government
of India; Swiss NSF; the Polish State Committee for Scientific
Research; STAA of Slovakia, and the Korea Sci. \& Eng. Foundation.


\begin{thebibliography}{}

\bibitem{star130auau} C.~Adler \textit{et al.}, Phys.~Rev.~Lett.~\textbf{89} 202301 (2002).

\bibitem{star200auau} J.~Adams, \textit{et al.}, Phys.~Rev.~Lett.~\textbf{91} 172302 (2003).

\bibitem{star200dau} J.~Adams, \textit{et al.}, Phys.~Rev.~Lett.~\textbf{91} 072304 (2003)

\bibitem{gluonrad} R.~Baier \textit{et al.}, Ann.~Rev.~Nucl.~Part.~Sci.~50, 37 (2000);
M.~Gyulassy \textit{et al.}, nucl-th/0302077.

\bibitem{pQCD} S.~Frixione \textit{et al.}, Adv. Ser. Direct. High Energy Phys. 15 609 (1998).

\bibitem{pQCDCalc1} R.~Vogt, hep-ph/0205330.

\bibitem{pQCDCalc2} M.~Cacciari \textit{et al.}, Phys.~Rev.~Lett.~\textbf{95} 122001 (2005);
FONLL calculations with CTEQ6M, $m_c = 1.5$ \gevcc, $m_b = 5$ \gevcc, and $\mu_{R,F} = m_T$.

\bibitem{Frixione:2005yf} S.~Frixione, Eur.\ Phys.\ J.\ C {\bf 43} 103 (2005).

\bibitem{STARDMesons1} J.~Adams, \textit{et al.}, Phys.~Rev.~Lett.~\textbf{94} 062301 (2005).

\bibitem{PhenixPP} A. Adare, \textit{et al.}, Phys. Rev. Lett.~{\bf 97}, 252002
(2006).

\bibitem{dead} Yu.~L.~Dokshitzer and D.E.~Kharzeev, Phys.~Lett.~B \textbf{519} 199 (2001).

\bibitem{dead1} B.~W.~Zhang \textit{et al.}, Phys.~Rev.~Lett.~\textbf{93} 072301 (2004).

\bibitem{elastic} M.~G.~Mustafa, Phys.~Rev.~C \textbf{72} 014905 (2005).

\bibitem{Djordjevic:2005} M.~Djordjevic \textit{et al.}, Phys. Lett. {\bf B632}
 81 (2006).

\bibitem{Armesto} N. Armesto \textit{et al.}, Phys.Lett. B~\textbf{637} 362 (2006).

\bibitem{Wicks:2005} S. Wicks \textit{et al.}, nucl-th/0512076.

\bibitem{Hess:2005} H. van Hess, V. Greco and R. Rapp, Phys. Rev. C~\textbf{73} 034913 (2006)
and private comunication.

\bibitem{Adler:2005xv}  S.~S.~Adler {\it et al.},  Phys.~Rev.~Lett.~\textbf{96} (2006) 032301.

\bibitem{Ackermann:2002ad} K.~H.~Ackermann {\it et al.}, Nucl.~Inst.~Meth.~A {\bf 499}
624 (2003).

\bibitem{refTPC}  M.~Anderson {\it et al.}, Nucl.~Inst.~Meth.~A {\bf 499}
659 (2003).

\bibitem{Beddo:2002zx}  M.~Beddo {\it et al.}, Nucl.~Inst.~Meth.~A {\bf 499} 725 (2003).

\bibitem{Shao} M.~Shao {\it et al.}, Nucl.~Instrum.~Meth.~A {\bf 558} 419 (2006).

\bibitem{Bichsel} H. Bichsel, Nucl. Instrum. Meth. B {\bf 562} 154 (2006).

\bibitem{Dalitz}  S. Eidelman {\it et al.}, Phys. Lett. B \textbf{592} 1 (2004).

\bibitem{embedding} J. Adams {\it et al.}, nucl-ex/0311017.

\bibitem{PHENIXpi0} S.S. Adler {\it et al.}, nucl-ex/0610036.

\bibitem{Sjostrand:2003wg} T.~Sjostrand \textit{et al.}, hep-ph/0308153.

\bibitem{Gyulassy:1994ew}  M.~Gyulassy and X.~N.~Wang, Comput.~Phys.~Commun.~{\bf 83} 307 (1994).

\bibitem{RAMONA} R. Vogt, private communication.

\bibitem{Magdalena} M. Djordjevic, nucl-th/0603066.

\bibitem{Liu} W.~Liu and C.~M.~Ko, nucl-th/0603004.

\end{thebibliography}
\end{document}